\documentclass[conference]{IEEEtran}
\usepackage{amssymb}
\usepackage{amsmath}
\usepackage{cite}
\usepackage{url}
\usepackage{xcolor}
\usepackage{cite,graphicx,amsmath,amssymb}
\usepackage{subfigure}
\usepackage{citesort}
\usepackage{fancyhdr}
\usepackage{mdwmath}
\usepackage{mdwtab}
\usepackage{caption}
\usepackage{amsthm}
\usepackage{algorithm}
\usepackage{algorithmic}
\allowdisplaybreaks
\usepackage[
top    = 0.75 in,
bottom = 1.05 in,
left   = 0.63 in,
right  = 0.63 in]{geometry}

\newtheorem{theorem}{Theorem}

\newtheorem{lemma}{Lemma}

\newtheorem{corollary}{Corollary}

\newcommand{\bm}[1]{\mbox{\boldmath{$#1$}}}
\begin{document}

\title{Intelligent Reflecting Surface Enhanced Indoor Robot Path Planning Using Radio Maps}
%

\author{\IEEEauthorblockN{Xidong Mu\IEEEauthorrefmark{1},Yuanwei Liu\IEEEauthorrefmark{2},
Li Guo\IEEEauthorrefmark{1}, Jiaru Lin\IEEEauthorrefmark{1} and Robert Schober\IEEEauthorrefmark{3}}
\IEEEauthorblockA{\IEEEauthorrefmark{1}Beijing University of Posts and Telecommunications, Key Lab of Universal Wireless Communications,\\ Ministry of Education, Beijing, China.\\ \IEEEauthorrefmark{2}Queen Mary University of London, London, UK.\\\IEEEauthorrefmark{3}Institute for Digital Communications, Friedrich-Alexander-University Erlangen-N{\"u}rnberg (FAU), Germany.\\E-mail:\{muxidong, guoli, jrlin\}@bupt.edu.cn, yuanwei.liu@qmul.ac.uk, robert.schober@fau.de}}

\maketitle
\begin{abstract}
An indoor robot navigation system is investigated, where an intelligent reflecting surface (IRS) is employed to enhance the connectivity between the access point (AP) and a mobile robotic user. The considered system is optimized for minimization of the travelling time/distance of the mobile robotic user from a given starting point to a predefined final location, while satisfying constraints on the communication quality. To tackle this problem, a \emph{radio map} based approach is proposed to exploit location-dependent channel propagation knowledge. Specifically, a \emph{channel power gain map} is constructed, which characterizes the spatial distribution of the maximum expected effective channel power gain of the mobile robotic user for the optimal IRS phase shifts. Based on the obtained channel power gain map, the communication-aware robot path planing problem is solved as a shortest path problem by exploiting graph theory. Numerical results show that: 1) Deploying an IRS can significantly extend the coverage of the AP and reduce the travelling distance of the mobile robotic user; 2) 2- or 3-bit IRS phase shifters can achieve nearly the same performance as continuous IRS phase shifters.
\end{abstract}
\section{Introduction}
\begin{figure}[t!]
    \begin{center}
        \includegraphics[width=2.6in]{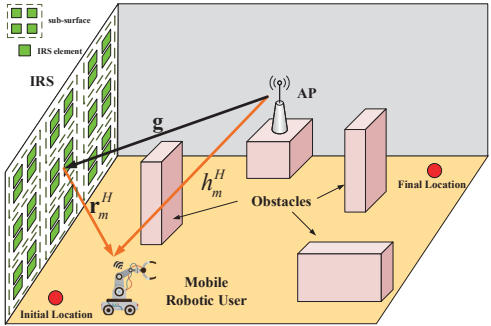}
        \caption{Illustration of the IRS-enhanced indoor robot navigation system for a single-user scenario.}
        \label{System model}
    \end{center}
\end{figure}

In the past few decades, robot technology has developed rapidly and has had a significant impact on human life \cite{Handbook_robot}. Specifically, robots can help humans perform repetitive or dangerous tasks, thus liberating human resources and reducing health risks. There is a wide range of robot applications, including cargo/packet delivery, search and rescue, public safety surveillance, environmental monitoring, and automatic industrial production \cite{robot_cloud,Robots_EM}. With the rapid development of fifth-generation (5G) and beyond (B5G) cellular networks, one promising solution is to integrate robots into cellular networks as connected robotic users to be served by base stations (BSs) or access points (APs). Connected robots can carry out missions relying on information exchange with operators \cite{connected_robot}, thus requiring less memory and computational resources compared to conventional robots which carry out tasks autonomously. Therefore, connected robots are more cost-efficient and less computation-constrained. Given the ultra-high speed, low latency, and high reliability of 5G/B5G networks, connected robots are expected to use the service of these networks in the future.\\
\indent Despite the appealing advantages of connected robots, one crucial limitation is that their communication link may be severely blocked by buildings, trees or other tall objects. The resulting signal \emph{dead zones} can significantly restrict the area of operation and reduce the efficiency of connected robots. Fortunately, with the recent advances in meta-materials, intelligent reflecting surfaces (IRSs) \cite{WuTowards,RIS_survey} have been proposed as an effective solution for overcoming signal blockage and enhancing the communication quality. An IRS is a thin man-made surface consisting of a large number of low-cost and passive reflecting elements (e.g., PIN diodes), each of which can reflect and impact the propagation of an incident electromagnetic wave \cite{WuTowards}. As a result, IRSs can create a \emph{programmable wireless environment}. For instance, if the signal transmission via the direct link is blocked, an IRS can be deployed to provide an additional reflected link, hence improving the received signal strength.\\
\indent Motivated by the aforementioned advantages, the performance gain facilitated by IRSs in wireless communication systems has been extensively investigated. The authors of \cite{Wu2019IRS} proposed an alternating optimization based algorithm for the design of the active beamforming at the BS and the passive beamforming at the IRS with the objective of minimizing the transmit power. The authors of \cite{Huang_EE} investigated energy-efficiency maximization in an IRS-assisted multiple-user multiple-input single-output (MISO) system. In \cite{Yu_secure}, the authors studied the physical layer security in IRS-aided communication systems, where the system sum secrecy rate was maximized. Furthermore, the authors of \cite{Huang_IRS_DL} invoked deep reinforcement learning techniques to tackle the joint active and passive beamforming problem.\\
\indent As mentioned earlier, signal blockage is the major bottleneck for connected robots. Motivated by this issue, we propose to deploy an IRS to assist the communication of an AP with a connected robot. In particular, an IRS-enhanced indoor robot navigation system is investigated, where one mobile robotic user is served by an AP with the aid of an IRS, see Fig. \ref{System model}. The mobile robotic user is dispatched to travel from a predefined initial location to a final location to carry out a specific mission. We aim to minimize the time/distance needed by the mobile robotic user to arrive at the final location by jointly optimizing the robot path and the IRS reflection matrix, while achieving a certain communication quality. Although the performance gain introduced by IRSs has been studied for various wireless communication system architectures, to the best of the authors' knowledge, this is the first work to investigate the IRS-enhanced indoor robot path planning problem. The main related challenges are as follows:
\vspace{-0.1cm}
\begin{itemize}
  \item As the signal transmission may be blocked by obstacles, as illustrated in Fig. 1, the channel power gain changes abruptly as the mobile robotic user travels. As a result, the location-dependent channel power gain makes the communication-aware robot path planning problem challenging.
  \item In addition, the channel power gain of the robotic user does not only depend on its location but also on the IRS phase shifts, which causes the path planning and the IRS reflection matrix design to be highly coupled.
\end{itemize}
\vspace{-0.1cm}
To overcome the aforementioned challenges, we develop a new radio-map based approach for the robot path planning problem. In general, a radio map contains information on the spectral activities and the propagation conditions in the space, frequency, and time domains \cite{radiomap}. This information can be exploited to improve the performance of wireless networks and facilitates new wireless applications. Inspired by this, we construct a specific type of radio map, namely, a \emph{channel power gain map}, by exploiting knowledge about the channel propagation conditions. The channel power gain map characterizes the spatial distribution of the maximum expected effective channel power gain of the mobile robotic user exploiting the IRS. Leveraging this map, the robot path planning problem is efficiently solved as a shortest path problem using graph theory. Numerical results show that the proposed IRS-enhanced system can significantly improve the channel power gain and reduce the travelling distance of the mobile robotic user compared to the conventional system without IRS.\\
\indent \emph{Notations:} Scalars, vectors, and matrices are denoted by lower-case, bold-face lower-case, and bold-face upper-case letters, respectively. ${\mathbb{C}^{N \times 1}}$ denotes the space of $N \times 1$ complex-valued vectors. The transpose and conjugate transpose of vector ${\mathbf{a}}$ are denoted by ${{\mathbf{a}}^T}$ and ${{\mathbf{a}}^H}$, respectively. ${\left\| {\mathbf{a}} \right\|_1}$ and $\left\| {\mathbf{a}} \right\|$ denote the 1-norm and the Euclidean norm of a vector ${\mathbf{a}}$, respectively.  ${\rm {diag}}\left( \mathbf{a} \right)$ denotes a diagonal matrix with the elements of vector ${\mathbf{a}}$ on the main diagonal. ${\left[ \cdot  \right]_n}$ denotes the $n$th element of a vector.
\section{System Model and Problem Formulation}
We consider an IRS-enhanced indoor robot navigation system, which consists of one single-antenna AP, one single-antenna mobile robotic user, and one IRS with $M$ passive reflecting elements, see Fig. \ref{System model}. The IRS is deployed on one of the indoor walls for assisting the transmission from the AP to the robotic user. Adopting a three-dimensional (3D) Cartesian coordinate system, the locations of the AP and the IRS are denoted by ${\mathbf{b}}=\left( {x_b,y_b,{H_b}} \right)$ and ${\mathbf{f}}=\left( {x_f,y_f,{H_f}} \right)$, respectively. The mobile robotic user is dispatched to travel from an initial location ${{\mathbf{q}}_I} = \left( {{x_I},{y_I},{H_0}} \right)$ to a final location ${{\mathbf{q}}_F} = \left( {{x_F},{y_F},{H_0}} \right)$, where ${{H_0}}$ denotes the height of the antenna of the mobile robotic user. Let ${\mathbf{q}}\left( t \right) = \left( {x\left( t \right),y\left( t \right),{H_0}} \right),t \in \left[ {0,T} \right]$, denote the time-varying path of the mobile robotic user, where $T$ denotes the required travelling time\footnote{The considered setup is representative for many practical connected robot applications, such as transportation of material in smart factories or delivery of medicine in hospitals.}. For practical implementation, the IRS is equipped with a smart controller, realized, e.g., with a field-programmable gate array (FPGA), which allows the AP to configure the IRS phase shifts in a real time manner. As the AP-IRS-user link suffers from severe path loss, a large number of reflecting elements are required for this link to achieve a comparable path loss as the unobstructed  direct AP-user link \cite{tile}. However, a large number of reflecting elements also cause a prohibitively high overhead/complexity for channel acquisition and phase shift design/reconfiguration. To overcome this limitation, an effective method is to group adjacent reflecting elements, which are expected to experience high channel correlation, together to a sub-surface, as was done in \cite{Yang_IRS}. All elements belonging to the same sub-surface are assumed to share the same reflection coefficient. In this paper, the $M$ passive reflecting elements of the IRS are divided into $N$ sub-surfaces, where each sub-surface consists of $\overline N  = {M \mathord{\left/
 {\vphantom {M N}} \right.
 \kern-\nulldelimiterspace} N}$ reflecting elements. An example where $\overline N =4$ elements are grouped into a sub-surface is illustrated in Fig. \ref{System model}. The instantaneous IRS reflection matrix is denoted by ${\mathbf{\Theta }}\left( t \right) = {\rm{diag}}\left( {{\bm{\theta }}\left( t \right) \otimes {{\mathbf{1}}_{\overline N \times 1 }}} \right)$, where ${\bm{\theta }}\left( t \right) = {\left[ {{\beta _1}\left( t \right){e^{j{\theta _1}\left( t \right)}},{\beta _2}\left( t \right){e^{j{\theta _2}\left( t \right)}}, \ldots ,{\beta _N}\left( t \right){e^{j{\theta _N}\left( t \right)}}} \right]^T}$, and ${{\theta _n}\left( t \right)}$ and ${{\beta _n}\left( t \right)}$ denote the instantaneous phase shift and attenuation coefficient of the $n$th sub-surface of the IRS, respectively. In this paper, we assume ${\theta _n}\left( t \right) \in \left[ {0,2\pi } \right)$ and ${\beta _n}\left( t \right) = 1,\forall n \in {\mathcal{N}},t \in {{\mathcal{T}}}$, where ${\mathcal{N}} = \left\{ {1, \ldots ,N} \right\}$ and ${{\mathcal{T}}} = \left[ {0,T} \right]$.\\
\indent We focus our attention on the downlink transmission from the AP to the mobile robotic user. The channel between the AP and the IRS is denoted by ${\mathbf{g}} \in {\mathbb{C}^{N \times 1}}$, and follows the Rician channel model. Hence, ${\mathbf{g}}$ can be expressed as
\vspace{-0.2cm}
\begin{align}\label{AP-IRS}
{\mathbf{g}} = \frac{{\sqrt {{{\mathcal{L}}_{AI}}} }}{{\sqrt {{K_{AI}} + 1} }}\left( {\sqrt {{K_{AI}}} \overline {\mathbf{g}}  + \widehat {\mathbf{g}}} \right),
\end{align}
\vspace{-0.5cm}

\noindent where ${{{\mathcal{L}}_{AI}}}$ is the distance-dependent path loss of the AP-IRS channel, $\overline {\mathbf{g}}$ denotes the deterministic line-of-sight (LoS) component, $\widehat {\mathbf{g}}$ denotes the random non-LoS (NLoS) component, which follows the Rayleigh distribution, and $K_{AI}$ is the Rician factor.\\
\indent Furthermore, let ${h}\left( {{\mathbf{q}}\left( t \right)} \right)\in {\mathbb{C}^{1 \times 1}}$ and ${{\mathbf{r}}}\left( {{\mathbf{q}}\left( t \right)} \right)\in {\mathbb{C}^{N \times 1}}$ denote the AP-mobile robotic user and IRS-mobile robotic user channels for mobile robotic user location ${\mathbf{q}}\left( t \right)$. We have
\vspace{-0.2cm}
\begin{align}\label{AP-robotic}
{h}\!\left( {{\mathbf{q}}\left( t \right)} \right)\!\! =\!\! \frac{{\sqrt {{\!{\mathcal{L}}_{AM}}\!\left( {{\mathbf{q}}\left( t \right)} \right)} }}{{\sqrt {\!{K_{AM}}\!\left( {{\mathbf{q}}\left( t \right)} \right)\! + \!1} }}\!\left(\! \!{\sqrt {\!{K_{AM}}\!\left( {{\mathbf{q}}\left( t \right)} \right)} {{\overline h}}\!\left( {{\mathbf{q}}\left( t \right)} \right) \!+\! {{\widehat h}}} \right)\!,
\end{align}
\vspace{-0.5cm}
\begin{align}\label{IRS-robotic}
{{\mathbf{r}}}\!\left( {{\mathbf{q}}\left( t \right)} \right)\!\! = \!\!\frac{{\sqrt {{{\mathcal{L}}_{IM}}\!\left( {{\mathbf{q}}\left( t \right)} \right)} }}{{\sqrt {{K_{IM}}\!\left( {{\mathbf{q}}\left( t \right)} \right)\! + \!1} }}\!\left(\!\! {\sqrt {{K_{IM}}\!\left( {{\mathbf{q}}\left( t \right)} \right)} {{\overline {\mathbf{r}} }}\!\left( {{\mathbf{q}}\left( t \right)} \right)\! + \!{{\widehat {\mathbf{r}}}}} \right)\!,
\end{align}
\vspace{-0.4cm}

\noindent where ${{{\mathcal{L}}_{AM}}\left( {{\mathbf{q}}\left( t \right)} \right)}$ and ${{{\mathcal{L}}_{IM}}\left( {{\mathbf{q}}\left( t \right)} \right)}$ denote the corresponding path losses. ${{{\widehat h}}\left( {{\mathbf{q}}\left( t \right)} \right)}$ and ${{{\overline {\mathbf{r}} }}\left( {{\mathbf{q}}\left( t \right)} \right)}$ are the location-dependent LoS components. ${{\widehat h}}$ and ${\widehat {\mathbf{r}}}$ denote the random Rayleigh distributed NLoS components. ${{K_{AM}}\left( {{\mathbf{q}}\left( t \right)} \right)}$ and ${{K_{IM}}\left( {{\mathbf{q}}\left( t \right)} \right)}$ denote the location-dependent Rician factors. For instance, if the signal transmission between the mobile robotic user at location ${{\mathbf{q}}\left( t \right)}$ and the AP/IRS is blocked by obstacles, the corresponding channel is classified as NLoS and we have ${K_{AM/IM}}\left( {{\mathbf{q}}\left( t \right)} \right) = 0$. Otherwise, it is classified as an LoS dominated channel and ${K_{AM/IM}}\left( {{\mathbf{q}}\left( t \right)} \right) = {\kappa _{AM/IM}}$, where ${\kappa _{AM/IM}}$ is a constant. \\
\indent Due to the high path loss, similar to \cite{Wu2019IRS}, signals that are reflected by the IRS two or more times are ignored. Therefore, the IRS-aided effective channel between the AP and the mobile robotic user can be expressed as
\vspace{-0.2cm}
\begin{align}\label{effective static}
{c}\left( t \right) = h^H\left( {{\mathbf{q}}\left( t \right)} \right) + {\mathbf{r}}^H\left( {{\mathbf{q}}\left( t \right)} \right){\mathbf{\Theta}} \left( t \right){\mathbf{g}}.
\end{align}
\vspace{-0.6cm}

\noindent We note that ${c}\left( t \right)$ is a random variable since it depends on random variables $\left\{ {\widehat {\mathbf{g}},{{\widehat {\mathbf{r}}}},{{\widehat h}}} \right\}$. In this paper, we are interested in the expected/average effective channel power gain, defined as ${\mathbb{E}}\left[ {{{\left| {{c}\left( t \right)} \right|}^2}} \right]$. A closed-form expression for ${\mathbb{E}}\left[ {{{\left| {{c}\left( t \right)} \right|}^2}} \right]$ is provided in the following lemma.
\vspace{-0.2cm}
\begin{lemma}\label{expected effective channel power gain}
\emph{The expected effective channel power gain of the mobile robotic user is given by}
\vspace{-0.2cm}
\begin{align}\label{expected robotic channel gain}
\begin{gathered}
  {\mathbb{E}}\left[ {{{\left| {{c}\left( t \right)} \right|}^2}} \right] \triangleq {\lambda }\left( t \right) =\! {\left| {\widetilde h^H\!\left(\! {{\mathbf{q}}\!\left( t \right)} \!\right)\! +\! {{\widetilde {\mathbf{r}}}^H}\!\left(\! {{\mathbf{q}}\!\left( t \right)} \!\right)\!{\mathbf{\Theta}} \left( t \right)\!\widetilde {\mathbf{g}}} \right|^2}\!\hfill \\
  \!+ {\eta _{IM}}\left(\! {{\mathbf{q}}\!\left( t \right)} \right) \!+\! {\eta _{AI}}{\eta _{IM}}\left(\! {{\mathbf{q}}\!\left( t \right)} \right)\left( {{K_{IM}}\left(\! {{\mathbf{q}}\!\left( t \right)} \right) \!+\! {K_{AI}} \!+\! 1} \right)M, \hfill \\
\end{gathered}
\end{align}
\vspace{-0.2cm}

\noindent \emph{where} ${\eta _{AM}}\left(\! {{\mathbf{q}}\!\left( t \right)} \right) = \frac{{{{{\mathcal{L}}}_{AM}}\left(\! {{\mathbf{q}}\!\left( t \right)} \right)}}{{{K_{AM}}\left(\! {{\mathbf{q}}\!\left( t \right)} \right) + 1}}$, ${\eta _{IM}}\left(\! {{\mathbf{q}}\!\left( t \right)} \right) = \frac{{{{{\mathcal{L}}}_{IM}}\left(\! {{\mathbf{q}}\!\left( t \right)} \right)}}{{{K_{IM}}\left(\! {{\mathbf{q}}\!\left( t \right)} \right) + 1}}$, ${\eta _{AI}} = \frac{{{{{\mathcal{L}}}_{AI}}}}{{{K_{AI}} + 1}}$, $\widetilde h^H\left(\! {{\mathbf{q}}\left( t \right)} \!\right) = \sqrt {{{{{{\eta}_{AM}}\left(\! {{\mathbf{q}}\left( t \right)} \!\right)}{K_{AM}}\left(\! {{\mathbf{q}}\left( t \right)} \!\right)}}} \overline h^H{\left(\! {{\mathbf{q}}\left( t \right)} \!\right)},$ ${\widetilde {\mathbf{r}}^H}\left(\! {{\mathbf{q}}\left( t \right)} \!\right) = \sqrt {{{{{{\eta}_{IM}}\left(\! {{\mathbf{q}}\left( t \right)} \!\right)}{K_{IM}}\left(\! {{\mathbf{q}}\left( t \right)} \!\right)}}} {\overline {\mathbf{r}}^H}\left(\! {{\mathbf{q}}\left( t \right)} \!\right),$ \emph{and} $\widetilde {\mathbf{g}} = \sqrt {{{{{\eta}}_{AI}}{{K_{AI}}}}} \overline {\mathbf{g}}.$
\begin{proof}
\emph{See Appendix~A.}
\end{proof}
\end{lemma}
\indent Let ${\mathbf{w}}^H\left( {{\mathbf{q}}\left( t \right)} \right) = \widetilde {\mathbf{r}}^H\left( {{\mathbf{q}}\left( t \right)} \right){\rm{diag}}\left( {\widetilde {{\mathbf{g}}}} \right) \in {\mathbb{C}^{1 \times M}}$ denote the cascaded LoS channel of the AP-IRS-mobile robotic user link before the reconfiguration of the IRS. Then, the corresponding combined composite channel associated with the $n$th sub-surface is given by ${\left[ {{{\widetilde {\mathbf{w}}}^H}\left( {{\mathbf{q}}\left( t \right)} \right)} \right]_n} = \sum\nolimits_{\overline n = 1}^{\overline N} {{{\left[ {{{\mathbf{w}}^H}\left( {{\mathbf{q}}\left( t \right)} \right)} \right]}_{\overline n + \left( {n - 1} \right)\overline N}}} ,\forall n \in {{\mathcal{N}}}$~\cite{Yang_IRS}. Therefore, the first term in ${\lambda}\left( t \right)$ can be rewritten as
\vspace{-0.2cm}
\begin{align}\label{expected robotic channel gain0}
  {\left| {\widetilde h^H\!\!\left(\! {{\mathbf{q}}\!\left(\! t \!\right)\!} \right)\! \!+\! {\mathbf{w}}^H\!\left(\! {{\mathbf{q}}\!\left(\! t \!\right)\!} \right)\!\!\left(\! {{\bm{\theta }}\!\left(\! t \!\right)\! \otimes\! {{\mathbf{1}}_{\overline N \!\times\!1  }}} \!\right)\!} \right|^2}\!\!\!= \!\!{\left| {\widetilde h^H\!\!\left(\! {{\mathbf{q}}\!\left(\! t \!\right)\!} \right)\! \!+\! \widetilde {\mathbf{w}}^H\!\left(\! {{\mathbf{q}}\!\left(\! t \!\right)\!} \right)\!{\bm{\theta }}\!\left(\! t \!\right)\!} \right|^2}\!\!.
\end{align}
\vspace{-0.4cm}

\indent  We aim to minimize the required travelling time $T$ of the mobile robotic user from ${{\mathbf{q}}_I}$ to ${{\mathbf{q}}_F}$ by jointly optimizing the path of the mobile robotic user, ${{Q}} = \left\{ {{\mathbf{q}}\left( t \right),0 \le t \le T} \right\}$, and the reflection matrix of the IRS $A = \left\{ {{\mathbf{\Theta}} \left( t \right),0 \le t \le T} \right\}$, subject to a constraint on the expected effective channel power gain. Hence, the communication-aware robot path planning problem can be formulated as
\vspace{-0.2cm}
\begin{subequations}\label{P_single}
\begin{align}
&\mathop {\min }\limits_{Q,A,T} \;T\\
\label{single user rate requirment2}{\rm{s.t.}}\;\;&{\lambda }\left( t \right) \ge \overline \gamma  ,\forall t \in {{\mathcal{T}}},\\
\label{IRS phase}&{\theta _n}\left( t \right) \in \left[ {0,2\pi } \right),\forall n \in {\mathcal{N}},t \in {{\mathcal{T}}},\\
\label{Initial Location Constraint}&{{\mathbf{q}}}\left( 0 \right) = {\mathbf{q}}_I,{{\mathbf{q}}}\left( T \right) = {\mathbf{q}}_F,\\
\label{speed}&\left\| {{\mathbf{\dot q}}\left( t \right)} \right\| \le {V_{\max }},\forall t \in {{\mathcal{T}}},
\end{align}
\end{subequations}
where the first derivative of ${{\mathbf{q}}}\left( t \right)$ with respect to $t$, ${{\mathbf{\dot q}}\left( t \right)}$, denotes the velocity vector, and $\overline \gamma$ denotes the minimum required expected effective channel power gain, which has to be achieved throughout the travel of the mobile robotic user. Constraints \eqref{Initial Location Constraint} and \eqref{speed} represent the mobility constraints on the mobile robotic user, where ${V_{\max }}$ is the maximum travelling speed. Problem \eqref{P_single} is challenging to solve for the following three reasons. Firstly, constraint \eqref{single user rate requirment2} is not concave with respect to ${\mathbf{q}}\left( t \right)$ and ${\mathbf{\Theta}}  \left( t \right)$. The unit modulus constraint \eqref{IRS phase} is also non-convex. Secondly, the expected effective channel power gain ${\lambda }\left( t \right)$ is generally not a continuous function under the considered location-dependent channel model. Thirdly, problem \eqref{P_single} involves an infinite number of optimization variables with respect to continuous time $t$, which are difficult to handle. To tackle these difficulties, we develop a radio map based approach which is capable of exploiting knowledge regarding location-dependent channel propagation.
\section{Radio Map based Approach}
In this section, we introduce a specific type of radio map, namely, the channel power gain map. Specifically, the channel power gain map characterizes the spatial distribution of the expected effective channel power gain over the region of interest with respect to the mobile robotic user's location ${\mathbf{q}}$, i.e., ${\lambda }\left( {\mathbf{q}} \right)$.
\subsection{Channel Power Gain Map Construction}
For the development of the radio map, the continuous two-dimensional (2D) space is first discretized into $\frac{{\overline X }}{\Delta }\frac{{\overline Y }}{\Delta }$ small cells, where $\Delta $ denotes the size of each cell and ${\overline X }$ and ${\overline Y }$ denote the range of the 2D space along the x-axis and y-axis, respectively. $\Delta $ should be chosen small enough such that the location of the mobile robotic user within each cell can be approximated by the cell center. For all $i \in {\mathcal{X}},j \in {\mathcal{Y}}$, the horizontal location of the $\left( {i,j} \right)$-th cell center can be expressed as
\vspace{-0.2cm}
\begin{align}\label{location u}
{\mathbf{q}}_{i,j}^\Delta  = {{\mathbf{q}}_0} + \left[ {i - 1,j - 1} \right]\Delta,
\end{align}
\vspace{-0.5cm}

\noindent where ${{\mathbf{q}}_0}$ is the center of the cell in the lower left corner of the considered 2D space, ${\mathcal{X}} = \left\{ {1, \ldots ,X} \right\}$, ${\mathcal{Y}} = \left\{ {1, \ldots ,Y} \right\}$, $X \triangleq \frac{{\overline X }}{\Delta }$, and $Y \triangleq \frac{{\overline Y }}{\Delta }$.\\
\indent Accordingly, let matrix ${\mathbf{C}} \in {{\mathbb{R}}^{X \times Y}}$ denote the channel power gain map, where the element in row $i$ and column $j$ characterizes the maximum expected effective channel power gain of the mobile robotic user at location $\left\{ {{\mathbf{q}}_{i,j}^\Delta } \right\}$. Therefore, the elements of ${\mathbf{C}}$ are given by
\vspace{-0.2cm}
\begin{align}\label{channel gain map}
{\left[ {\mathbf{C}} \right]_{i,j}} = \mathop {\max }\limits_{{\mathbf{\Theta }} \in {{\mathcal{F}}}} {\left| {\widetilde h^H\left( {{\mathbf{q}}_{i,j}^\Delta } \right) + \widetilde {\mathbf{w}}^H\left( {{\mathbf{q}}_{i,j}^\Delta } \right){\bm{\theta }}} \right|^2} + {\tau }\left( {{\mathbf{q}}_{i,j}^\Delta } \right),
\end{align}
\vspace{-0.5cm}

\noindent where $i \in {\mathcal{X}},j \in {\mathcal{Y}}$, ${\mathcal{F}}$ denotes the set of all possible IRS reflection matrices, and ${\tau}\left( {{\mathbf{q}}_{i,j}^\Delta } \right) = {\eta _{AM}}\left( {{\mathbf{q}}_{i,j}^\Delta } \right) + {\eta _{AI}}{\eta _{IM}}\left( {{\mathbf{q}}_{i,j}^\Delta } \right)\left( {{K_{IM}}\left( {{\mathbf{q}}_{i,j}^\Delta } \right) + {K_{AI}} + 1} \right)N$.\\
\indent For any given ${\mathbf{q}}_{i,j}^\Delta$, the expected effective channel power gain is upper-bounded by
\vspace{-0.2cm}
\begin{align}\label{upper bound}
\begin{gathered}
  {\left| {\widetilde h^H\left( {{\mathbf{q}}_{i,j}^\Delta } \right) + \widetilde {\mathbf{w}}^H\left( {{\mathbf{q}}_{i,j}^\Delta } \right){\bm{\theta }}} \right|^2} + {\tau }\left( {{\mathbf{q}}_{i,j}^\Delta } \right) \hfill \\
   \le {\left( {\left| {\widetilde h^H\left( {{\mathbf{q}}_{i,j}^\Delta } \right)} \right| + {{\left\| {\widetilde {\mathbf{w}}^H\left( {{\mathbf{q}}_{i,j}^\Delta } \right)} \right\|}_1}} \right)^2} + {\tau }\left( {{\mathbf{q}}_{i,j}^\Delta } \right).\hfill \\
\end{gathered}
\end{align}
\vspace{-0.4cm}

\noindent The above inequality holds with equality for the following optimal phase shifts:
\vspace{-0.2cm}
\begin{align}\label{ap theta}
\theta _n^*\left(\! {{\mathbf{q}}_{i,j}^\Delta } \!\right) \!=\! \angle \left(\! {\widetilde h^H{{\left(\! {{\mathbf{q}}_{i,j}^\Delta } \!\right)}}} \!\right)\! -\! \angle \left(\! {{{\left[ {\widetilde {\mathbf{w}}^H\left(\! {{\mathbf{q}}_{i,j}^\Delta } \!\right)} \right]}_n}} \!\right),\forall n\! \in \!{{\mathcal{N}}},
\end{align}
\vspace{-0.5cm}

\noindent Therefore, for all $i \in {\mathcal{X}},j \in {\mathcal{Y}}$, the channel power gain map ${\mathbf{C}}$ is given as follows:
\vspace{-0.2cm}
\begin{align}\label{obtained channel gain map}
  {\left[ {\mathbf{C}} \right]_{i,j}} = {\left( {\left| {\widetilde h^H\left( {{\mathbf{q}}_{i,j}^\Delta } \right)} \right| + {{\left\| {\widetilde {\mathbf{w}}^H\left( {{\mathbf{q}}_{i,j}^\Delta } \right)} \right\|}_1}} \right)^2} + {\tau }\left( {{\mathbf{q}}_{i,j}^\Delta } \right).
\end{align}
\vspace{-0.8cm}
\subsection{Optimal Path}
Let ${{Q}} = \left\{ {{\mathbf{q}}_{{i_1},{j_1}}^\Delta ,{\mathbf{q}}_{{i_2},{j_2}}^\Delta , \ldots ,{\mathbf{q}}_{{i_{D - 1}},{j_{D - 1}}}^\Delta ,{\mathbf{q}}_{{i_D},{j_D}}^\Delta } \right\}$ denote the path of the mobile robotic user. For ease of exposition, we assume that ${\mathbf{q}}_{{i_1},{j_1}}^\Delta  = {{\mathbf{q}}_I}$ and ${\mathbf{q}}_{{i_D},{j_D}}^\Delta  = {{\mathbf{q}}_F}$. It can be verified that for the optimal solution of \eqref{P_single}, the speed constraint \eqref{speed} must be satisfied with equality, i.e., $\left\| {{\mathbf{\dot q}}\left( t \right)} \right\| = {V_{\max }},\forall t \in {{\mathcal{T}}}$. To demonstrate this, suppose that at the optimal solution to problem \eqref{P_single}, the mobile robotic user travels at a speed strictly less than ${V_{\max }}$. Then, we can increase the speed to ${V_{\max }}$, which decreases the travelling time. With this insight, problem \eqref{P_single} can be equivalently reformulated as the following \emph{travelling distance minimization} problem over the channel power gain map:
\vspace{-0.2cm}
\begin{subequations}\label{graph single}
\begin{align}
&\mathop {\min }\limits_{{{Q}},D} \sum\limits_{n = 1}^{D - 1} {\left\| {{\mathbf{q}}_{{i_{d + 1}},{j_{d + 1}}}^\Delta  - {\mathbf{q}}_{{i_d},{j_d}}^\Delta } \right\|}\\
\label{graph single QoS}{\rm{s.t.}}\;\;& {\left[ {\mathbf{C}} \right]_{{i_d},{j_d}}} \ge {\overline \gamma} ,\\
\label{adjacent single}&\left\| {{\mathbf{q}}_{{i_{d + 1}},{j_{d + 1}}}^\Delta  - {\mathbf{q}}_{{i_d},{j_d}}^\Delta } \right\| \le \sqrt 2 \Delta ,1 \le d \le D - 1,\\
\label{graph single Initial Location}&{\mathbf{q}}_{{i_1},{j_1}}^\Delta  = {{\mathbf{q}}_I},{\mathbf{q}}_{{i_D},{j_D}}^\Delta  = {{\mathbf{q}}_F},
\end{align}
\end{subequations}
\vspace{-0.5cm}

\noindent where \eqref{adjacent single} ensures that any two successive waypoints along the path are adjacent in the channel power gain map. This constraint ensures that if the two successive waypoints satisfy the expected effective channel power gain condition, then any point on the line segment between them also satisfies this condition. However, problem \eqref{graph single} is a non-convex combinatorial optimization problem, which is difficult to solve with standard convex optimization methods. In the following, we solve problem \eqref{graph single} by exploiting graph theory \cite{Graph}.\\
\indent For given ${\overline \gamma}$ and channel power gain map ${\mathbf{C}}$, we construct a new matrix ${\mathbf{\Pi}}  \in {{\mathbb{R}}^{X \times Y}}$, namely the feasible map, as follows:
\vspace{-0.6cm}
\begin{align}\label{feasible map}
{\left[ {\mathbf{\Pi}}  \right]_{i,j}}  = \left\{ \begin{gathered}
  1,\;\;{\rm{if}}\;{\left[ {\mathbf{C}} \right]_{i,j}} \ge \;{\overline \gamma}  \hfill \\
  0,\;{\rm{otherwise}} \hfill \\
\end{gathered}  \right.,i \in {\mathcal{X}},j \in {\mathcal{Y}}.
\end{align}
\vspace{-0.4cm}

\noindent Specifically, ${\left[ {\mathbf{\Pi}}  \right]_{i,j}} = 1$ means that the location ${{\mathbf{q}}_{i,j}^\Delta }$ is a feasible candidate waypoint for the path of the mobile robotic user.\\
\indent Based on the feasible map ${\mathbf{\Pi}}$, we construct an undirected weighted graph, which is denoted by $G = \left( {V,E} \right)$. The vertex set $V$ and the edge set $E$ are given by
\vspace{-0.2cm}
\begin{subequations}
\begin{align}\label{V}&V = \left\{ {{{\mathbf{v}}_{i,j}} = {\mathbf{q}}_{i,j}^\Delta :{{\left[ {\mathbf{\Pi}}  \right]}_{i,j}} = 1,i \in {\mathcal{X}},j \in {\mathcal{Y}}} \right\},\\
\label{E}&E = \left\{ {\left( {{{\mathbf{v}}_{i,j}},{{\mathbf{v}}_{i',j'}}} \right):{{\mathbf{v}}_{i,j}},{{\mathbf{v}}_{i',j'}} \in V} \right\}.
\end{align}
\end{subequations}
\vspace{-0.5cm}

\noindent The weight of each edge is denoted by $W\left( {{{\mathbf{v}}_{i,j}},{{\mathbf{v}}_{i',j'}}} \right)$, and given by
\vspace{-0.2cm}
\begin{align}
W\!\!\left( {{{\mathbf{v}}_{i,j}},\!{{\mathbf{v}}_{i',j'}}} \right)\!\! =\!\! \left\{ \begin{gathered}
  \!\!\left\| {{{\mathbf{v}}_{i,j}} \!- \!{{\mathbf{v}}_{i',j'}}} \right\|\!,{\rm{if}}\!\left\| {{{\mathbf{v}}_{i,j}}\! -\! {{\mathbf{v}}_{i',j'}}} \right\| \!\le \!\sqrt 2 \Delta  \hfill \\
  \!\!\infty ,{\rm{otherwise}} \hfill \\
\end{gathered}.  \right.
\end{align}
\vspace{-0.3cm}

\noindent Based on the constructed graph $G$, problem \eqref{graph single} is equivalent to finding the shortest path from ${\mathbf{v}}_{{i_1},{j_1}} = {{\mathbf{q}}_I}$ to ${\mathbf{v}}_{{i_D},{j_D}} = {{\mathbf{q}}_F}$. The shortest path construction problem can be efficiently solved via the Dijkstra algorithm \cite{Graph} with complexity ${\mathcal{O}}\left( {{{\left| V \right|}^2}} \right)$. The optimal path for the mobile robotic user is denoted by ${{{Q}}^*} = \left\{ {{\mathbf{q}}_{{i_1},{j_1}}^{*\Delta },{\mathbf{q}}_{{i_2},{j_2}}^{*\Delta }, \ldots ,{\mathbf{q}}_{{i_{D - 1}},{j_{D - 1}}}^{*\Delta },{\mathbf{q}}_{{i_D},{j_D}}^{*\Delta }} \right\}$.
\section{Numerical Examples}
\begin{figure*}[htb!]
\centering
\begin{minipage}[t]{0.3\linewidth}
\includegraphics[width=2.2in]{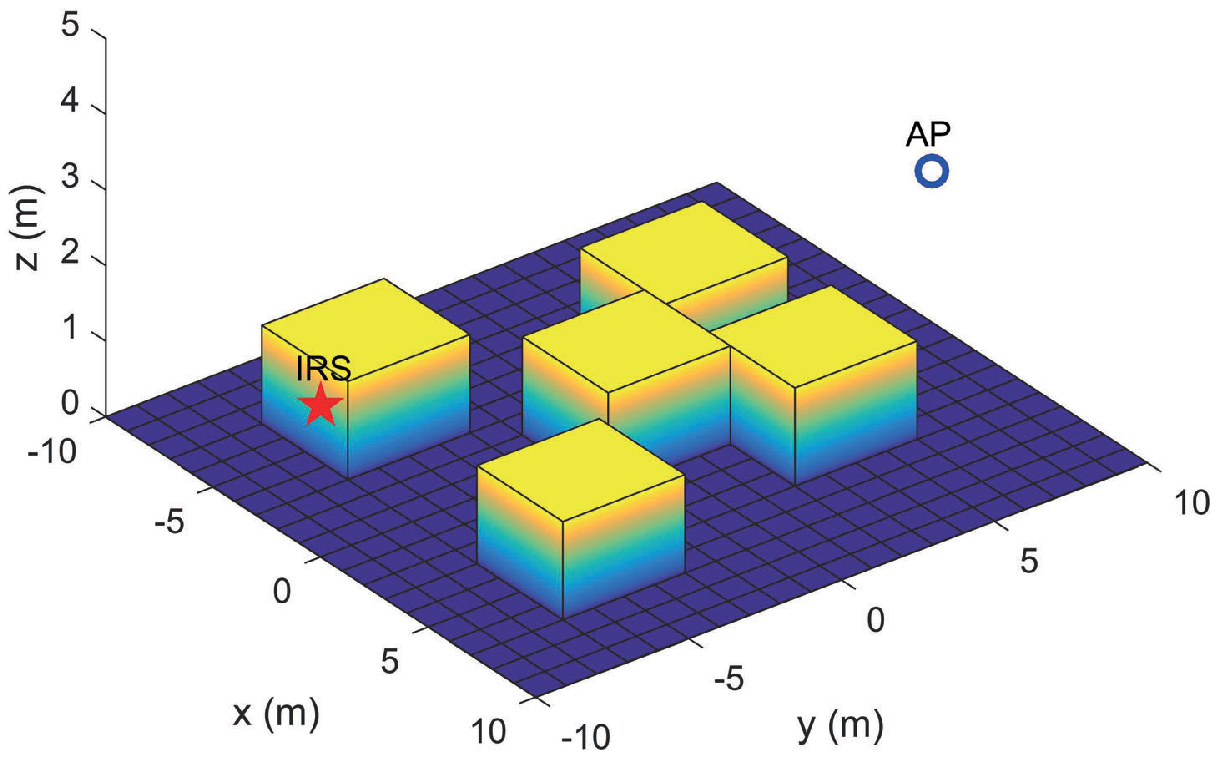}
\caption{The simulated scenario (3D view).}
\label{setup}
\end{minipage}
\quad
\begin{minipage}[t]{0.6\linewidth}
\subfigure[Without IRS]{\label{withoutIRS}
\includegraphics[width= 2.2in]{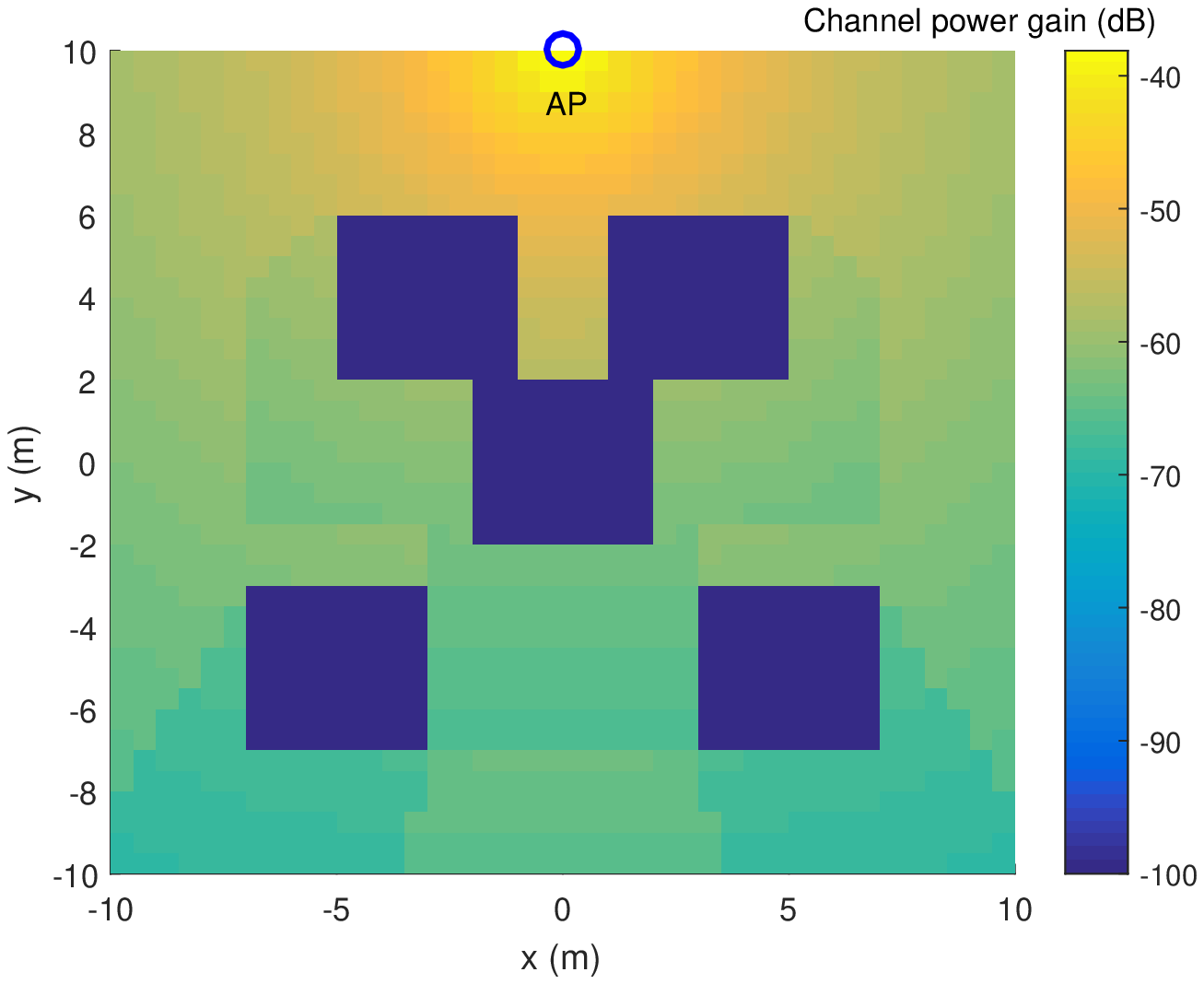}}
\subfigure[With IRS]{\label{withIRS}
\includegraphics[width= 2.2in]{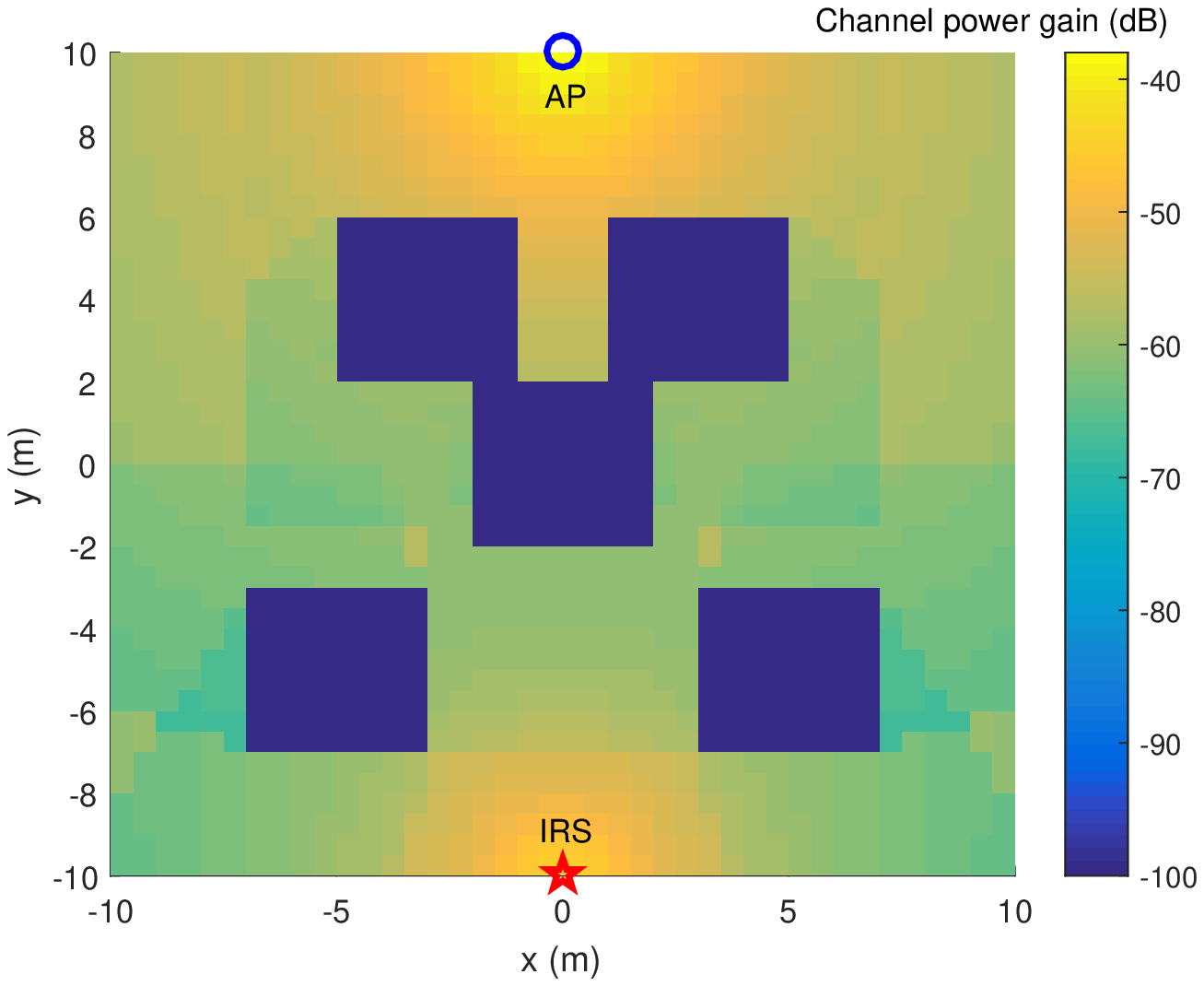}}
\setlength{\abovecaptionskip}{-0cm}
\caption{Illustration of the obtained channel power gain map for different schemes and ${\Delta }=0.5$ m.}\label{channel gain map1}
\label{DvS}
\end{minipage}
\end{figure*}
In this section, numerical examples are provided to validate the performance of the proposed IRS-enhanced robot navigation system. As illustrated in Fig. \ref{setup}, we consider an indoor factory (InF) environment with a width and length of 20 meter, respectively, and a ceiling height of 5 meter. Specifically, the AP and the IRS are deployed at $\left( {0,10,2} \right)$ meters and $\left( {0,-10,2} \right)$ meters, respectively. The number of IRS reflecting elements in each sub-surface is set to $\overline N  = 20$. The total number of sub-surfaces is $N = {N_x}{N_z}$, where ${N_x}$ and ${N_z}$ denote the number of sub-surfaces along the $x$-axis and $z$-axis, respectively. Therefore, the total number of IRS reflecting elements is $M = \overline N {N_x}{N_z}$, where we set ${N_x}=10$  and increase ${N_z}$ linearly with $M$. The considered indoor environment includes 5 obstacles with a size of $4 \times 4 \times 1.3$ ${\rm{m}}^3$, respectively. The horizontal centers of the obstacles are located at $\left( { - 5, - 5} \right)$, $\left( { 5, - 5} \right)$, $\left( { 0, 0} \right)$, $\left( { - 3, 4} \right)$, and $\left( { 3, 4} \right)$ meters. The height of the antenna of the mobile robotic user is ${H_0} = 1$ m and its initial and final locations are ${{\mathbf{q}}_I} = \left( { - 10,0,1} \right)$ meters and ${{\mathbf{q}}_F} = \left( { 10,0,1} \right)$ meters, respectively. The path losses of all involved channels are modeled according to the 3rd Generation Partnership Project (3GPP) technical report for the InF-SH (sparse clutter, high BS) scenario \cite{3GPP}. For LoS channels, the path loss in dB is given by
\vspace{-0.2cm}
\begin{align}
{{\mathcal{L}}}_{{\rm{LoS}}} = 31.84 + 21.50{\log _{10}}\left( d \right) + 19{\log _{10}}\left({f_c}\right),
\end{align}
\vspace{-0.6cm}

\noindent where $d$ denotes the 3D distance between the robotic user and the AP (or the IRS), and ${f_c} = 2$ GHz is the carrier frequency. For NLoS channels, the path loss in dB is given by
\vspace{-0.2cm}
\begin{align}
{{\mathcal{L}}_{{\rm{NLoS}}}}\! =\! \max \left\{ {{{\mathcal{L}}_{{\rm{LoS}}}},32.4\! +\! 23{{\log }_{10}}\!\left( d \right)\! +\! 20{{\log }_{10}}\!\left({f_c}\right)} \right\},
\end{align}
\vspace{-0.6cm}

\noindent which ensures that ${{\mathcal{L}}_{{\rm{NLoS}}}} \ge {{\mathcal{L}}_{{\rm{LoS}}}}$. The Rician factors of all involved channels are set to 3 dB.\\
\indent For comparison, we also consider the following benchmark schemes:
\begin{itemize}
  \item \textbf{IRS with discrete phase shifts}: In this case, the IRS is assumed to be equipped with finite resolution phase shifters. We have ${\theta _n} \in {\mathcal{D}} = \left\{ {0,\delta , \ldots ,\left( {L - 1} \right)\delta } \right\}$, where $\delta  = {{2\pi } \mathord{\left/
 {\vphantom {{2\pi } L}} \right.
 \kern-\nulldelimiterspace} L}$ and $L$ denotes the number of discrete phase shift levels. The corresponding channel power gain map is obtained by quantizing the optimal phase shift $\theta _n^* \left( {{\mathbf{q}}_{i,j}^\Delta } \right)$ in \eqref{ap theta} to the nearest discrete phase shift in ${\mathcal{D}}$ as follows:
 \vspace{-0.6cm}
 \begin{align}\label{ap1 theta}
{ \theta  _n^{\mathcal{D}}}\left( {{\mathbf{q}}_{i,j}^\Delta } \right) = \mathop {\arg \min  }\limits_{\theta  \in {\mathcal{D}}} \left| {\theta  - \theta _n^*\left( {{\mathbf{q}}_{i,j}^\Delta } \right)} \right|,\forall n \in {\mathcal{N}}.
\end{align}
\vspace{-0.4cm}
  \item \textbf{Without IRS}: In this case, the AP serves the user without the help of an IRS. The channel power gain map is obtained by considering only the AP-user channel.
\end{itemize}
\subsubsection{Channel Power Gain Map} Fig. \ref{channel gain map1} illustrates the channel power gain map obtained from \eqref{channel gain map} with and without IRS, respectively. We set the size of each cell to ${\Delta }=0.5$ m and the number of reflecting elements is $M=1200$. As the mobile robotic user cannot enter the regions covered by obstacles, the corresponding expected channel power gain is set to $ - \infty $. One can observe that the distribution of the channel power gain changes abruptly due to the obstacles. Specifically, as depicted in Fig. \ref{withoutIRS}, without IRS, the channel power gains severely degrade if the AP-user link is blocked by obstacles. Moreover, from Fig. \ref{withIRS}, it can be observed that the channel power gains can be considerably improved by deploying an IRS, especially for the cells around the IRS. The IRS can be interpreted as a virtual AP, however, it is more energy-efficient than an actual AP since the IRS only passively reflects the incident signals.
\subsubsection{Obtained Path of the Mobile Robotic User} 
\begin{figure*}[htb!]
\centering
\begin{minipage}[t]{0.3\linewidth}
\includegraphics[width=2.2in]{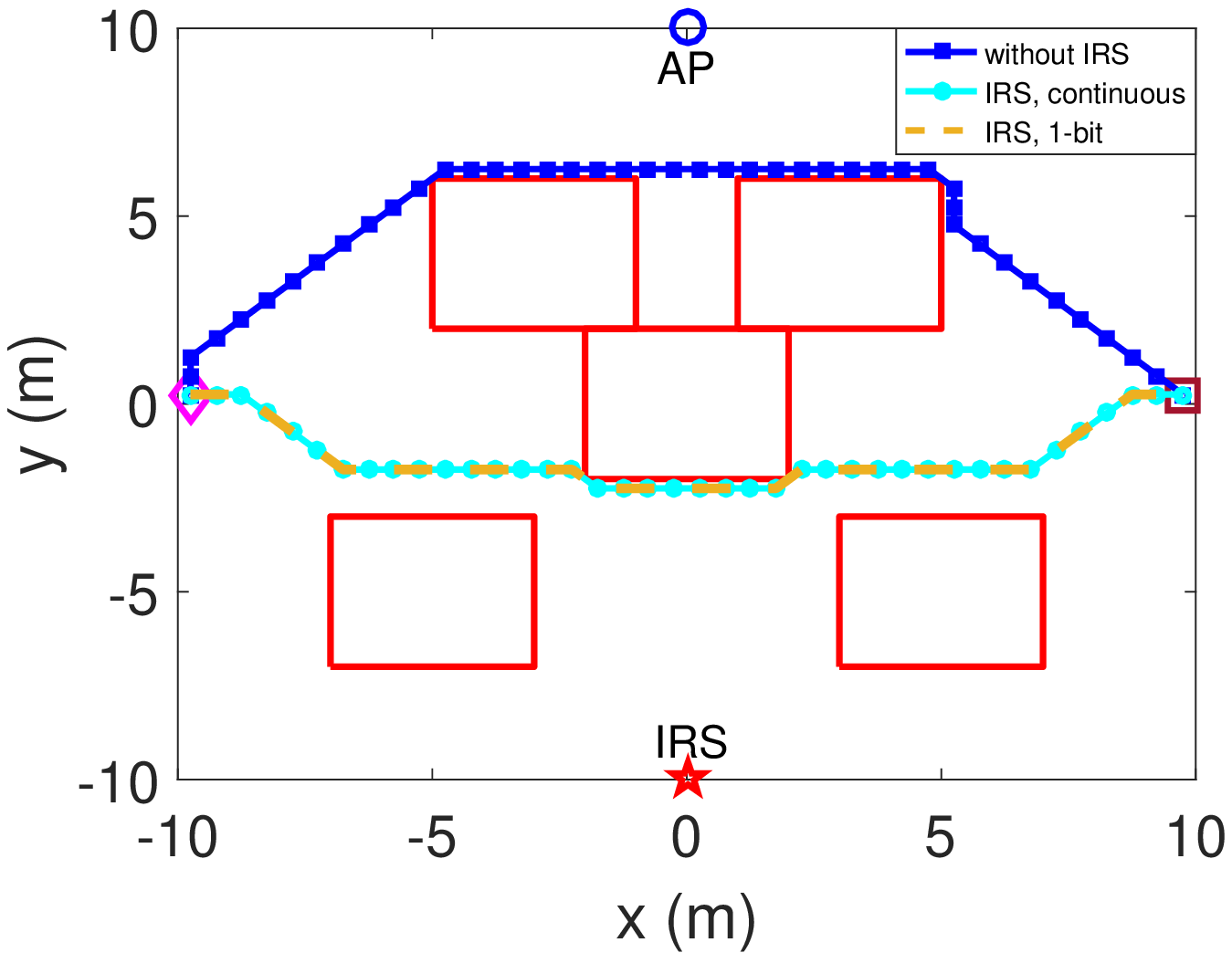}
\caption{The obtained paths of the mobile robotic user for $\overline \gamma = -63$ dB and $M=1200$.}
\label{db63}
\end{minipage}
\quad
\begin{minipage}[t]{0.3\linewidth}
\includegraphics[width=2.2in]{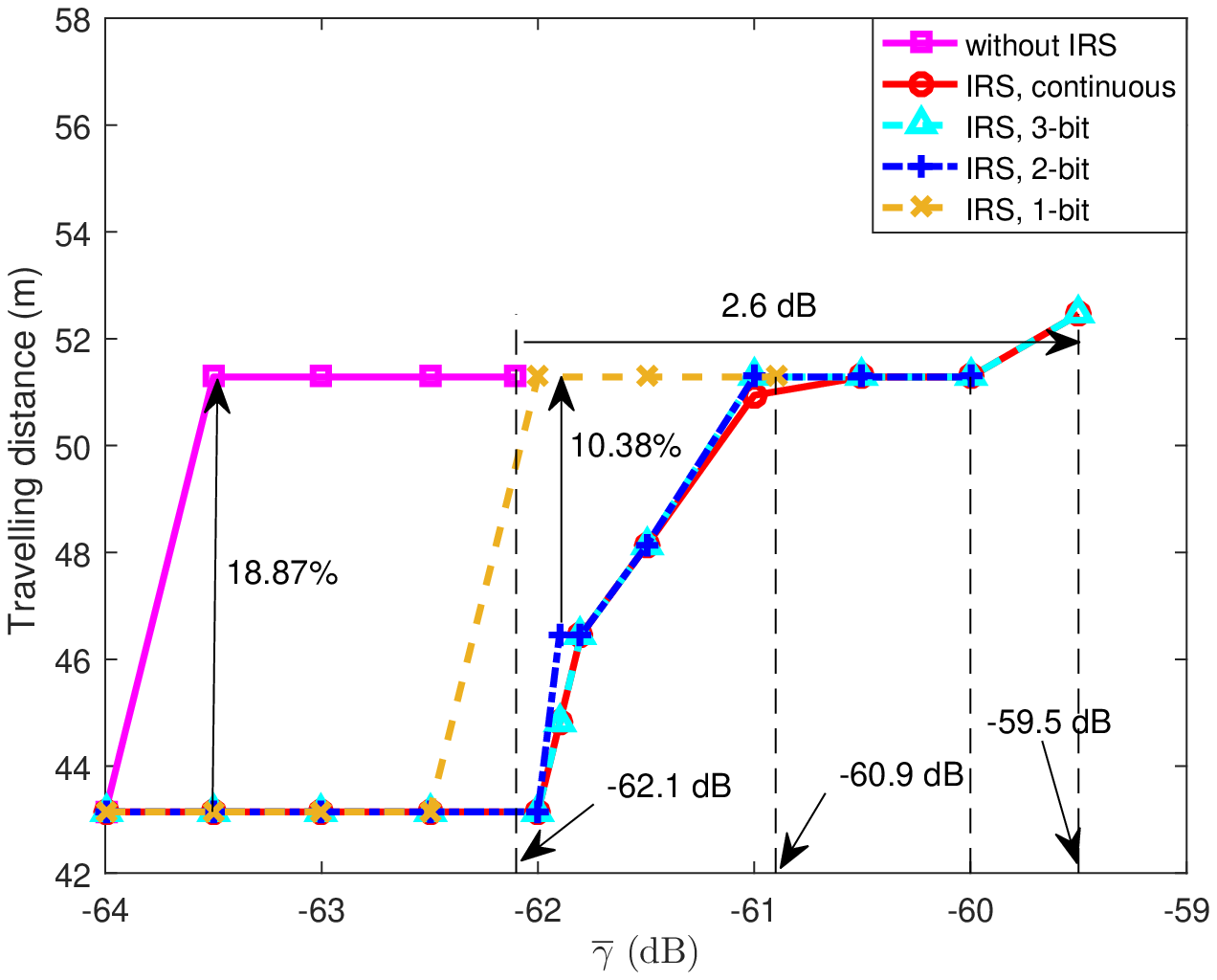}
\caption{Travelling distance versus $\overline \gamma$ for $M=1200$.}
\label{DvT}
\end{minipage}
\quad
\begin{minipage}[t]{0.3\linewidth}
\includegraphics[width=2.2in]{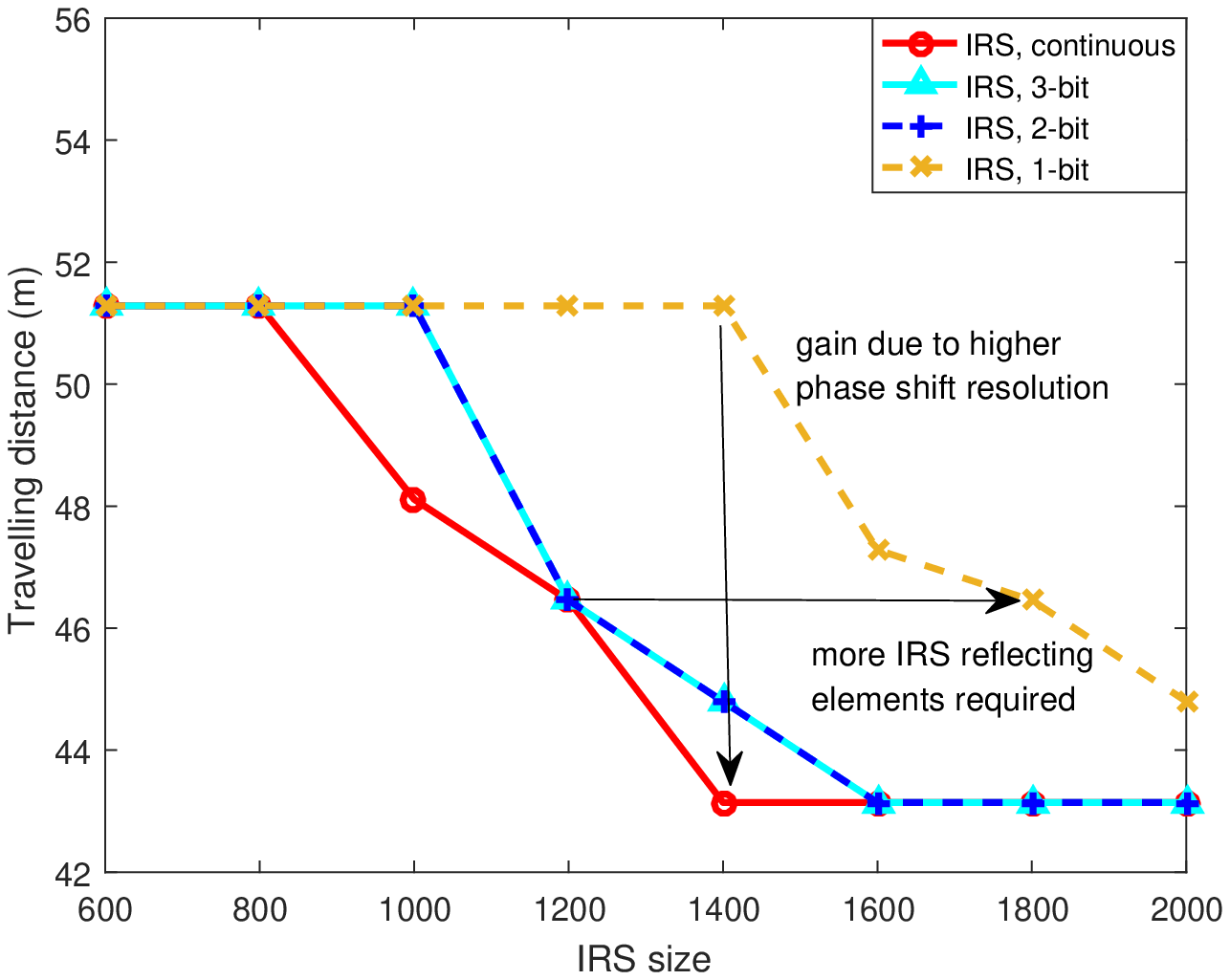}
\caption{Travelling distance versus $M$ for $\overline \gamma=-61.8$ dB.}%
\label{DvS}
\end{minipage}
\end{figure*}

Fig. \ref{db63} depicts the obtained paths of the mobile robotic user for different schemes. The red boxes represent the regions covered by obstacles. The initial and final locations of the mobile robotic user are denoted by ``$\lozenge$'' and ``$\square$'', respectively. For comparison, results without IRS and for 1-bit quantization are also shown. As can be observed in Fig. \ref{db63}, for $\overline \gamma = -63$ dB and $M=1200$, the path obtained for the case without IRS approaches the AP to avoid the blockage caused by the obstacles. This is expected since only travelling along such a path can create a good channel condition for the mobile robotic user, which in turn leads to a longer travelling distance. However, for the IRS-aided schemes, the mobile robotic user tends to travel in a relatively straight line from ${\mathbf{q}}_I$ to ${\mathbf{q}}_F$, which leads to a shorter travelling distance compared to the case without IRS. Though the communication link between the mobile robotic user and the AP may be blocked by obstacles, a reflected LoS dominated communication link can be established with the IRS. Therefore, the mobile robotic user is not forced to travel towards the AP, since the IRS offers more degrees of freedom for path planning. This clearly demonstrates the benefits of deploying an IRS.
\subsubsection{Travelling Distance versus Expected Channel Power Gain Target} In Fig. \ref{DvT}, we depict the travelling distance of different schemes versus the expected channel power gain target $\overline \gamma$ for $M=1200$. It is first observed that the minimum required travelling distances of all schemes generally increase as $\overline \gamma$ increases. This is expected since a larger expected channel power gain requirement reduces the number of feasible cells in the radio map, which also reduces the flexibility in path planning. Note that without IRS, the path planning problem becomes infeasible for $\overline \gamma \ge -62.1$ dB. The feasibility threshold for the IRS-aided schemes increases to $-62.1$ dB, $-60.9$ dB, $-60$ dB, and $-59.5$ dB as the phase shift resolution improves. The proposed scheme with continuous phase shifts yields a 2.6 dB performance gain over the scheme without IRS. Moreover, without IRS, when  $- 63.5\; {\rm{dB}}\le \overline \gamma   \le - 62.5$ dB the mobile robotic user needs to travel up to 18.87\% farther than when the IRS is present. Furthermore, with the 1-bit phase shifter, the required travelling distance is at most 10.38\% larger than for higher phase shift resolutions when $- 61.9\; {\rm{dB}}\le \overline \gamma   \le - 60.9$ dB. The performance degradation caused by 2- or 3-bit phase shifters is negligible compared to continuous phase shifters.
\subsubsection{Travelling Distance versus Number of IRS Elements} In Fig. \ref{DvS}, the required travelling distance for different schemes versus the number of IRS elements $M$ is presented. We set the expected channel power gain target to $\overline \gamma=-61.8$ dB. As can be observed, in general, the minimum required travelling distance of each scheme decreases as $M$ increases. This is because a larger number of IRS elements is capable of achieving a higher array gain, which allows the mobile robotic user to travel in a more flexible manner. Furthermore, to achieve the same travelling distance, the 1-bit phase shifter requires at most 600 additional IRS elements compared to the other phase shifters. The performance achieved by 2- or 3-bit phase shifters is close to that with continuous phase shifters. This reveals an interesting trade-off between the number of IRS elements and the number of phase shift resolution bits. Though a smaller number of phase shift resolution bits reduces the cost of the IRS elements, it increases the required number of IRS elements to achieve a certain performance, which in turn increases the deployment cost.
\vspace{-0.3cm}
\section{Conclusions}
\vspace{-0.2cm}
An IRS-assisted indoor robot navigation system has been investigated. The communication-aware robot path planning problem was formulated for minimization of the travelling time/distance by jointly optimizing the robot path and the phase shifts of the IRS elements. To solve this problem, we proposed a radio map based approach, where a channel power gain map depicting the spatial distribution of the maximum expected effective channel power gain of the mobile robotic user is constructed. Based on the channel power gain map, the robot path planning problem was efficiently solved by invoking graph theory. Numerical results showed that the coverage of the AP can be significantly extended by deploying an IRS, and the robot travelling distance can be significantly reduced with the aid of an IRS.
\vspace{-0.3cm}
\section*{Appendix~A: Proof of Lemma~\ref{expected effective channel power gain}} \label{Appendix:A}
\vspace{-0.2cm}
The expected effective channel power gain of the mobile robotic user, ${\mathbb{E}}\left[ {{{\left| {{c}\left( t \right)} \right|}^2}} \right]$, can be decomposed as follows:
\vspace{-0.3cm}
\begin{align}\label{x0}
\begin{gathered}
  {\mathbb{E}}\left[ {{{\left| {{c}\left( t \right)} \right|}^2}} \right] = {\mathbb{E}}\left\{ {{{\left| {{h}{{\left( {{\mathbf{q}}\left( t \right)} \right)}^H} + {{\mathbf{r}}}{{\left( {{\mathbf{q}}\left( t \right)} \right)}^H}{\mathbf{\Theta}} \left( t \right){\mathbf{g}}} \right|}^2}} \right\} \hfill \\
   = \!{\mathbb{E}}\left\{ {{{\left| {\left( {\widetilde h^H\!\left( {{\mathbf{q}}\left( t \right)} \right)\! +\! \breve h^H} \right) \!+\! \left( {\widetilde {\mathbf{r}}^H\left( {{\mathbf{q}}\left( t \right)} \right) \!+\! \breve {\mathbf{r}}^H} \right)\!{\mathbf{\Theta}} \left( t \right)\!\left( {\widetilde {\mathbf{g}} \!+\! \breve {\mathbf{g}}} \right)} \right|}^2}} \right\} \hfill \\
  \mathop  = \limits^{\left( a \right)}\! {\left| {{x_1}} \right|^2} \!+\! {\mathbb{E}}\left\{\! {{{\left| {{x_2}} \right|}^2}} \!\right\} \!+\! {\mathbb{E}}\left\{\! {{{\left| {{x_3}} \right|}^2}} \!\right\} \!+\! {\mathbb{E}}\left\{\! {{{\left| {{x_4}} \right|}^2}} \!\right\} \!+\! {\mathbb{E}}\left\{\! {{{\left| {{x_5}} \right|}^2}} \!\right\}, \hfill \\
\end{gathered}
\end{align}
\vspace{-0.3cm}

\noindent where $\widetilde h^H\left( {{\mathbf{q}}\left( t \right)} \right) \!=\! \sqrt {{{{{{\eta}_{AM}}\left( {{\mathbf{q}}\left( t \right)} \right)}{K_{AM}}\left( {{\mathbf{q}}\left( t \right)} \right)}}} \overline h^H{\left( {{\mathbf{q}}\left( t \right)} \right)}$, $\breve h^H \!=\! \sqrt {{{{{{\eta}}_{AM}}\left( {{\mathbf{q}}\left( t \right)} \right)}}} \widehat h^H$, ${\widetilde {\mathbf{r}}^H}\left( {{\mathbf{q}}\left( t \right)} \right) \!=\!\sqrt {{{{{\eta}}_{IM}}\left( {{\mathbf{q}}\left( t \right)} \right){{K_{IM}}\left( {{\mathbf{q}}\left( t \right)} \right)}}} {\overline {\mathbf{r}} ^H}\left( {{\mathbf{q}}\left( t \right)} \right)$, $\breve {\mathbf{r}}^H \!=\!\sqrt {{{{{{\eta}}_{IM}}\left( {{\mathbf{q}}\left( t \right)} \right)}}} {\widehat {\mathbf{r}}^H} $, $\widetilde {\mathbf{g}} \!=\! \sqrt {{{{{{\eta}}_{AI}}{K_{AI}}}}} \overline {\mathbf{g}} $, and $\breve {\mathbf{g}} \!=\! \sqrt {{{{{{\eta}}_{AI}}}}} \widehat {\mathbf{g}}$. In \eqref{x0}, $\left( a \right)$ is due to the fact that $\breve h^H$, $\breve {\mathbf{r}}^H$, and $\breve {\mathbf{g}}$ have zero means and are independent from each other. We have
\vspace{-0.2cm}
\begin{subequations}
\begin{align}
\label{x1}&{\left| {{x_1}} \right|^2} = {\left| {\widetilde h^H\left( {{\mathbf{q}}\left( t \right)} \right) + \widetilde {\mathbf{r}}^H\left( {{\mathbf{q}}\left( t \right)} \right){\mathbf{\Theta}} \left( t \right)\widetilde {\mathbf{g}}} \right|^2},\\
\label{x2}&{\mathbb{E}}\left\{ {{{\left| {{x_2}} \right|}^2}} \right\} = {\mathbb{E}}\left\{ {{{\left| \breve {h^H} \right|}^2}} \right\} = {{{\eta}_{AM}}\left( {{\mathbf{q}}\left( t \right)} \right)},\\
\label{x3}&{\mathbb{E}}\!\left\{ {{{\!\left| {{x_3}} \right|}^2}}\! \right\}\! \!=\! {\mathbb{E}}\!\left\{\! {{{\left| {\widetilde {\mathbf{r}}^H\!\!\left( \!{{\mathbf{q}}\!\left( t \right)}\! \right)\!{\mathbf{\Theta}}\! \left( t \right)\!\breve {\mathbf{g}}} \right|}^2}}\! \right\} \!\!=\! {\eta _{A\!I}}{\eta _{I\!M}}\!\left( \!{{\mathbf{q}}\!\left( t \right)}\! \right)\!{K_{I\!M}}\!\left( \!{{\mathbf{q}}\!\left( t \right)}\! \right)\!M,\\
\label{x4}&{\mathbb{E}}\left\{ {{{\left| {{x_4}} \right|}^2}} \right\} \!=\! {\mathbb{E}}\left\{ {{{\left| {\breve {\mathbf{r}}^H\!\!\left( \!{{\mathbf{q}}\!\left( t \right)}\! \right){\mathbf{\Theta}} \left( t \right)\widetilde {\mathbf{g}}} \right|}^2}} \right\}  \!=\! {\eta _{AI}}{K_{AI}}{\eta _{IM}}\left( \!{{\mathbf{q}}\!\left( t \right)}\! \right)\!M,\\
\label{x5}&{\mathbb{E}}\!\left\{\! {{{\left| {{x_5}} \right|}^2}\!} \right\} \!=\! {\mathbb{E}}\!\left\{ {{{\left| {\breve {\mathbf{r}}^H\!\!\left( \!{{\mathbf{q}}\!\left( t \right)}\! \right){\mathbf{\Theta}}\! \left( t \right)\breve {\mathbf{g}}} \right|}^2}} \right\}  \!=\! {\eta _{AI}}{\eta _{IM}}\left( \!{{\mathbf{q}}\!\left( t \right)}\! \right)\!M.
\end{align}
\end{subequations}
\vspace{-0.4cm}

\noindent Therefore, by inserting the results in \eqref{x1}-\eqref{x5} into \eqref{x0}, we arrive at \eqref{expected robotic channel gain}. This completes the proof of Lemma 1.
\bibliographystyle{IEEEtran}
\bibliography{mybib}
\end{document}